\documentclass[12pt]{article}

\usepackage{makeidx}
\usepackage{epsfig}
\usepackage{color}
\usepackage{graphicx}
\usepackage{xspace}
\makeindex

\usepackage{amsmath}
\usepackage{amsthm}
\usepackage{amscd}
\usepackage{amsfonts}
\usepackage[psamsfonts]{amssymb}

\def\Time{{t}}
\def\Space{{x}}

\begin{document}
\begin{titlepage}

\begin{flushright}
UCB-PTH-07/15\\
LBNL-63372
\end{flushright}

\vfil\

\begin{center}{\Large{\bf World Sheet Commuting $\beta\gamma$ CFT }} \\
{\Large{\bf and Non-Relativistic String Theories}}
\vfil

Bom Soo Kim

\bigskip

Department of Physics, University of California, Berkeley, CA 94720\\
Theoretical Physics Group, Lawrence Berkeley National Laboratory, Berkeley, CA 94720\\
{\it bskim@socrates.berkeley.edu}
\vfil

\end{center}

\begin{abstract}

We construct a sigma model in two dimensions with Galilean symmetry in flat target space 
similar to the sigma model of the critical string theory with Lorentz symmetry 
in 10 flat spacetime dimensions. 
This is motivated by the works of Gomis and Ooguri\cite{Gomis:2000bd} and 
Danielsson et. al.\cite{Danielsson:2000gi, Danielsson:2000mu}. 
Our theory is much simpler than their theory and does not assume a compact coordinate. 
This non-relativistic string theory has a bosonic matter $\beta\gamma$ CFT with the conformal 
weight of $\beta$ as 1. 
It is natural to identify time as a linear combination of $\gamma$ and $\bar{\gamma}$ through 
an explicit realization of the Galilean boost symmetry. 
The angle between $\gamma$ and $\bar{\gamma}$ parametrizes one parameter family 
of selection sectors. 
These selection sectors are responsible for having a non-relativistic dispersion relation without 
a nontrivial topology in the non-relativistic setup, which is one of the major differences 
from the previous works\cite{Gomis:2000bd, Danielsson:2000gi, Danielsson:2000mu}. 
This simple theory is the non-relativistic analogue of the critical string theory, and
there are many different avenues ahead to be investigated. 
We mention a possible consistent generalization of this theory with different 
conformal weights for the $\beta\gamma$ CFT. 
We also mention supersymmetric generalizations of these theories.

\end{abstract}

\vspace{0.5in}

\end{titlepage}

\section{Introduction}

String theory is a prime candidate for a quantum theory of gravity and is 
widely studied with the relativistic target space symmetry. A starting point at the perturbative level is 
the Polyakov action with flat target spacetime with Lorentz symmetry as its global symmetry. 
Open strings by themselves are not consistent as they can join end points together
and turn into closed strings. But recently it has been realized that, 
in a small part of its moduli space, string theory can 
have only open strings without closed strings\cite{Seiberg:2000ms,Gopakumar:2000na}.
These theories opened up new possibilities to study the nature of string theory without complication of gravity. 
It has been a fascinating subject by itself revealing many novel properties 
including a space and time noncommutativity.

Further studies revealed that it is also possible to have a closed string theory with a different 
target space symmetry. That target space symmetry is Galilean symmetry, and it was named 
Non-Relativistic Closed String Theory\cite{Gomis:2000bd}. Strikingly, this non-relativistic string theory has 
a world sheet description\cite{Gomis:2000bd} as well as a target space 
description\cite{Danielsson:2000gi, Danielsson:2000mu}.\footnote
{The non-relativistic nature is interesting from several aspects. Here is one:
whereas the complete fundamental description of string theory is yet to be clarified,
it has been put forward that a fully nonperturbative definition 
of noncritical M-theory in 2+1 dimensions can be written in terms of a non-relativistic 
Fermi liquid\cite{Horava:2005tt}. We will return to this point in the conclusion, where we 
try to justify a consideration of string thoery with the non-relativistic setup.}
The world sheet action is simple, but the analysis 
of these papers heavily relied on the original relativistic description with an 
assumption of a compact coordinate in order to preserve the effect of the NSNS B-field.
Despite its strong connection to the original theory, the world sheet description attracts much attention 
(see {\it e.g.}, \cite{Brugues:2004an}).
We will start to investigate this world sheet theory with some modifications of the action and 
of the target space topology. These modifications are considered, and are partially justified, in section 2.

The purpose of this paper is to propose a simpler world sheet action for the non-relativistic 
string theory of Gomis and Ooguri\cite{Gomis:2000bd} and of Danielsson et. al.
\cite{Danielsson:2000gi, Danielsson:2000mu}, and to study its properties.
In this paper we investigate a basic bosonic sigma model with flat spatial coordinates and 
with a matter $\beta\gamma$ CFT replacing time and one of the spatial coordinates of the Polyakov 
action, similarly to the CFT of \cite{Gomis:2000bd, Danielsson:2000gi, Danielsson:2000mu}. 
The main differences in our model are: (i) there is no compact coordinate in our description, 
and (ii) there are no terms other than the $\beta\gamma$ CFT action. 
On the way of developing this theory, we realize that it is possible to have a one-parameter family 
of selection sectors which parametrize the target-space time coordinate. Each sector in this family 
is represented by a different linear combination of $\gamma$ and $\bar{\gamma}$. We explicitly construct 
the Galilean boost transformation with this generalized time coordinate. 
We construct the general vertex operators following the work of Gomis and Ooguri\cite{Gomis:2000bd}. 
We propose a spacetime interpretation for the $\beta\gamma$ CFT, and we also find some restrictions 
on the parameters in the ground state vertex operator. We explicitly quantize the theory with the 
``Old Covariant Quantization'' scheme and also with BRST quantization. We calculate some correlation 
functions and demonstrate the consistency of the theory by checking the modular invariance. 
The $\beta\gamma$ zero modes play a crucial 
role in the spacetime interpretation. These are the contents of section 3.

We then proceed to generalize this theory by allowing
more general conformal weights for the $\beta\gamma$ fields (in the bosonic string theory).
If we want to have a target space interpretation, there is only a finite range of the allowed conformal  
weights of these fields. 
It seems that these theories are the non-relativistic analogues of the noncritical string theories. 
These are interesting because these theories can shed light on the (relativistic) non-critical string 
theories. We discuss these theories in section 4 and present some immediate observations.

We conclude in section 5, where we explain some implications of this bosonic non-relativistic string theory. 
We also mention a supersymmetric generalization of this theory. Some justifications for 
considering the string theory in the non-relativistic setup are also presented.

\section{Review and setup the starting point}

It is interesting to look briefly at some earlier developments regarding the low energy limit of open 
string theory associated with a large magnetic $B$ field (i.e., the spatial components of the NSNS 2-form field) 
or a critical electric $B$ field.
We therefore start with a review of Non-Commutative Open String (NCOS), 
Open Membrane (OM) and Non-Relativistic Closed String (NRCS) theories.
We further motivate the study of our non-relativistic string action. The ultimate justification, though, comes 
when we quantize and check the one loop consistency of our non-relativistic string theory at the end of section 3. 
If familiar with those theories, readers are encouraged to jump to the section 3.

Before we proceed, we recall that, in contrast to NCOS, the effective description of the low energy limit 
associated with a large value of the magnetic $B$ field is captured by noncommutative Yang-Mills theory with a 
space noncommutativity\cite{Seiberg:1999vs}.

A low energy limit of open string theory related to a critical electric $B$ field is not a field 
theory but a consistent open string theory formulated on a noncommutative spacetime (NCOS) with all the massive 
excitations of the open string in it\cite{Seiberg:2000ms, Gopakumar:2000na}. This can be understood by thinking 
about the open string as a dipole whose endpoints carry opposite charges\cite{Seiberg:2000ms}. Then, open strings 
stretched along the direction of the $B$ field are energetically favored. At the critical value of the electric 
$B$ field, the energy stored in tension is almost balanced by the electric energy of the stretched string, and the 
effective tension of the open string goes to $0$. In the low energy limit, we cannot ignore these light degrees 
of freedom, the open strings. On the other hand, strings on the brane can not turn into closed strings, because 
it will cost a lot of energy. Effectively, this theory is the theory of open strings, and the underlying spacetime 
is noncommutative (NCOS)\cite{Seiberg:2000ms}.   

It turns out that this phenomenon is more general and can be extended to the M5 brane theory with a critical 
electric 3-form field in M-theory. The tension of a membrane stretched along the directions of the electric 
3-form field is very light and cannot be ignored in the low energy limit. This theory is Open Membrane (OM) 
theory\cite{Gopakumar:2000ep}. And the S-dual of (5+1) dimensional NCOS theory in Type IIB is Open D1 brane 
theory on NS5 brane in Type IIB. With T-duality we can get Open Dp brane theories on the NS5 brane. 
These theories are a large class of 6 dimensional nongravitational theories with light open D branes 
among their excitations, which have near critical RR gauge fields of different ranks\cite{Gopakumar:2000ep}.

When the spatial coordinate along a critical electric $B$ field is compactified on a circle in NCOS theory, 
there are finite closed string states with the positive winding number which do not decouple from the open 
string spectrum\cite{Klebanov:2000pp}. From these observations the authors\cite{Gomis:2000bd}
\cite{Danielsson:2000gi} tried to understand the low energy limit of closed string theories with 
a compact coordinate in the presence of a background NSNS electric B field. They end up having a 
Non-Relativistic Closed String theory(NRCS) and II A/B Wound and Wrapped theories. For the latter case, 
neither a critical electric B-field nor D-branes were necessary to have non-relativistic 
dispersion relation in the low energy limit\cite{Danielsson:2000gi}. They also consider a low energy 
limit of critical RR fields and found Galilean invariant D-p brane solutions. It is useful to look at 
those a little further to motivate the current work.  

Gomis and Ooguri\cite{Gomis:2000bd} developed a world sheet description of the non-relativistic 
closed string theory by taking a consistent low energy limit of the relativistic
string theory. This limit is typically related to a critical value of an electric 
component of a background NSNS B field in order to cancel divergences which arise when 
one takes the low energy limit. A spatial coordinate along the electric B field should 
be compact to obtain nontrivial physical states. Otherwise the background B field can be 
gauged away without changing string spectra. Winding modes from this compact 
coordinate were important to obtain the non-relativistic energy dispersion relation,  
and the winding number multiplied by compactified radius had a role of mass. 
The resulting action of the low energy limit was written as 
\begin{eqnarray}
S_{GO} = \int \frac{d^2z}{2\pi} \left( \beta \bar{\partial} \gamma + \bar{\beta} \partial \bar{\gamma} 
+ \frac{1}{4 \alpha_{eff}'} \partial \gamma \bar{\partial} \bar{\gamma} + \frac{1}{\alpha_{eff}'} 
\partial X^i \bar{\partial} 
X_i\right), \label{goaction}
\end{eqnarray}
where $\gamma = X^0 + X^1$, $\bar{\gamma} = - X^0 + X^1$, and $\beta$, $\bar{\beta}$ are commuting auxiliary 
fields which were introduced as Lagrange multipliers 
through the process of taking the low energy limit.
The index $i$ of the fields $X^i$ runs from 2 to 9 and $\alpha_{eff}'$ is an effective string ``slope'' 
related to the compactification radius.  

Interestingly, at the same time Gomis and Ooguri published their paper, 
Danielsson et, al\cite{Danielsson:2000gi} provided a complementary description of this non-relativistic string theory. 
It was motivated by the observation that if there is a compact coordinate 
``NCOS'' D-strings can emit wound strings into the bulk\cite{Klebanov:2000pp}. 
Furthermore, through T-duality along the compact coordinate, it was possible to identify NCOS theory with the 
DLCQ description of the IIA string theory. The authors showed that one can define a meaningful 
`NCOS' limit of the IIA/B closed string theory, a theory of closed strings with positive winding modes, 
as long as there is a compact dimension. This theory provides a spacetime description of 
the string theory with the non-relativistic energy momentum relation, which is called ``Wound IIA/B'' theory.

From the world sheet formulation\cite{Gomis:2000bd} a 4-point scattering amplitude was 
calculated, and it revealed that there exist instantaneous Newtonian gravitational 
interactions whose origin could not be explained. Subsequently Danielsson 
et. al. further investigated this issue and provided the origin of these interactions as 
massless gravitons\cite{Danielsson:2000mu}. Actually massless gravitons are not dynamical degrees 
of freedom in the non-relativistic string point of view. They are 
sub-leading contributions from a zero winding sector of the non-relativistic string theory.
When Gomis and Ooguri derived the low energy limit, they had a term 
$- \int \frac{d^2 z}{2\pi} \left(\frac{2\alpha'}{1+2\pi\alpha' B} \beta\bar{\beta}\right)$ 
in their action which is responsible for the sub-leading contributions.
They then took the strict low energy limit which removed these sub-leading contributions. 
Even though this term was absent, the world sheet formulation was powerful 
enough to produce correct instantaneous gravitational interactions\cite{Gomis:2000bd}. 
Thus we will not consider a similar term in our action. 

The term $ \int \frac{d^2z}{2\pi} \left( \frac{1}{4 \alpha_{eff}'} \partial \gamma 
\bar{\partial} \bar{\gamma} \right)$
in the world sheet action (\ref{goaction}) can also be safely ignored 
without changing the physical spectrum of the non-relativistic string\cite{Danielsson:2000mu}. 
For the non-relativistic closed string spectrum, Danielsson 
et. al. showed that this term is just a leftover after removing a divergent 
contribution when one takes the low energy limit. 
For the non-relativistic open string spectrum, they explicitly showed 
that this term actually does not change the spectrum at all. 
Thus we will consider an action without a similar term. 
Gomis and Ooguri kept this term, which gives a constant contribution to the non-relativistic energy, 
in order to provide a precise connection of the non-relativistic 
string spectrum to the NCOS spectrum. 

These explain our starting point. If we follow the steps explained above, the remaining parts of the 
Gomis-Ooguri action are very simple and look very familiar. It is nothing but the conventional 
$\beta\gamma$ CFT with the conformal weights of $\beta$ and $\gamma$ as 1 and 0, respectively. 
Now it is time to start with the simple action with $\beta\gamma$ CFT and $X$ CFTs and 
to study its properties. While we proceed, we encounter many surprises.

\section{``Critical'' Non-relativistic Bosonic String Theory}

\noindent We start with a bosonic string theory action with a commuting $\beta\gamma$ CFT 
in addition to the spatial $X^i$ CFT, in conformal gauge.  
\begin{eqnarray}
S_0 = \int \frac{d^2z}{2\pi} \left( \beta \bar{\partial} \gamma + \bar{\beta} \partial \bar{\gamma} 
+  \frac{1}{\alpha'} \partial X^i \bar{\partial} X_i + b_g \bar{\partial} c_g 
+ \bar{b}_g \partial \bar{c}_g
\right), \label{bosonicaction1}
\end{eqnarray}
where $i$ runs from 1 to 24 for $X^i$ CFTs. The commuting matter $\beta\gamma$ CFT has 
conformal weights $h(\beta) = 1$ and $h(\gamma) = 0$. The central charge of the commuting 
$\beta\gamma$ CFT is $2$. The anticommuting ghost CFT, 
whose central charge is $-26$, has weights $h(b_g) = 2$ and $h(c_g) = -1$, as usual. 
To be anomaly free the total central charge of the whole system should be $0$, and we need 
$24$ spatial coordinates as indicated above. We will consider the cases with general 
$\lambda$ and corresponding $d$, and present a table for these theories in the next section. 
The case presented in this section, with $\lambda = 1$, is rather special, 
and we will refer to it as ``critical'' non-relativistic string theory.

We briefly comment on the commuting $\beta\gamma$ CFT\cite{fms}\cite{pol} with $\lambda = 1$. The OPEs of these fields are given by  
\begin{equation}
\gamma(z_1) \beta(z_2) \sim  \frac{1}{z_{12}}\ , ~~~~~ \beta(z_1) \gamma(z_2) \sim -\frac{1}{z_{12}}\ .
\end{equation}
The antiholomorphic fields satisfy similar OPEs.
The mode expansions and hermiticity properties are 
\begin{eqnarray}
\gamma(z) &=& \sum_{n=-\infty}^{\infty} \frac{\gamma_n}{z^{n}}  , ~~~~ \gamma_n^{\dagger} 
=  \gamma_{-n},    \\
\beta(z) &=& \sum_{n=-\infty}^{\infty} \frac{\beta_n}{z^{n+ 1}} , ~~~~ \beta_{n}^{\dagger} 
=  - \beta_{-n}. 
\end{eqnarray}
The holomorphic energy momentum tensor and its mode expansion are 
\begin{equation}
T(z) = \beta \partial \gamma = \sum_{n = -\infty}^{\infty}\frac{L_n}{z^{n+2}} , ~~~
L_n = \sum_{m=-\infty}^{\infty} (n - m) :\gamma_{n-m}\beta_{m}: .
\end{equation}
Importantly, the normal-ordering constant turns out to be $0$ for this critical case with $\lambda = 1$.

\subsection{Galilean Invariance and Selection Sectors}

Galilean invariance of the non-relativistic world sheet action was pointed out first in\cite{Gomis:2000bd} 
and written explicitly in \cite{Danielsson:2000mu} with a particular time coordinate given by 
$\Time_{GO} = \frac{1}{\sqrt{2}}(\gamma - \bar{\gamma})$. Actually, there exists a generalized Galilean invariance in the action (\ref{bosonicaction1}) which has generalized time 
$\Time = p\gamma(z) + q\bar{\gamma}(\bar{z}) = \cos(\phi) \gamma(z) + \sin(\phi) \bar{\gamma}(\bar{z})$, 
where $p = \cos(\phi)$ and $q = \sin(\phi)$.  
The Galilean boost transformation can be written in the following way 
\begin{eqnarray}
X^i &\longrightarrow& X^i + \frac{v^i}{2} \sqrt{\alpha'}~ (p\gamma(z) + q\bar{\gamma}(\bar{z})),  \nonumber\\
\beta &\longrightarrow& \beta - \frac{v^i}{\sqrt{\alpha'}}~ p~ \partial X^i - \frac{v^i v_i}{4}~ p~ 
\partial (p\gamma(z) + q\bar{\gamma}(\bar{z})), \\
\bar{\beta} &\longrightarrow& \bar{\beta} - \frac{v^i}{\sqrt{\alpha'}} ~q~ \bar{\partial} 
X^i - \frac{v^i v_i}{4}~ q~ \bar{\partial} (p\gamma(z) + q\bar{\gamma}(\bar{z})), \nonumber
\end{eqnarray}
where $v^i$ is the Galilean boost parameter. We can easily show that the above action is 
invariant upto total derivatives under this generalized Galilean boost transformation. 
And because the fields $\gamma$ and $\bar{\gamma}$ do not transform at all, there exist 
an infinite number of ``selection sectors'' parametrized by a ``selection time'' 
$\Time = \cos(\phi) \gamma(z) + \sin(\phi) \bar{\gamma} (\bar{z})$ for the non-relativistic 
string theory. As long as one belongs to one specific sector one will never escape 
from that sector with combinations of Galilean transformations. Non-relativistic 
closed string theory\cite{Gomis:2000bd} and Wound string theory\cite{Danielsson:2000gi}\cite{Danielsson:2000mu} 
deal with a special selection sector with $\Time_{GO} = \frac{1}{\sqrt{2}}(\gamma - \bar{\gamma})$ 
which corresponds to the case $\phi = (2n - \frac{1}{4})\pi$ for any integer $n$.

In addition to this Galilean boost invariance, the action (\ref{bosonicaction1}) is also 
invariant under $SO(8)$ rotations acting on the spatial coordinates $X^i$, and under overall 
spacetime translations given by, 
\begin{eqnarray}
X^i \rightarrow X^i + a^i, ~~~~~~~~~ \gamma \rightarrow \gamma + a^\gamma, ~~~~~~~~~ 
\bar{\gamma}
 \rightarrow \bar{\gamma} + a^{\bar{\gamma}},
\end{eqnarray}
where $a^i$, $a^\gamma$ and $a^{\bar{\gamma}}$ are constants.

\subsection{Vertex Operators}


Following Gomis and Ooguri\cite{Gomis:2000bd} we will start with a ground state vertex operator $V_0$ 
acting on the $SL(2,C)$ invariant vacuum.
\begin{eqnarray}
V_{0}(k^\gamma, k^{\bar{\gamma}}, k^i; z, \bar{z}) = g: e^{ik^\gamma \gamma + ik^{\bar{\gamma}} 
\bar{\gamma} - ip' \int^{z} \beta - iq' \int^{\bar{z}} \bar{\beta} + ik^i \cdot X_i} : , 
\label{vertexoperator}
\end{eqnarray}
where $k^\gamma$, $k^{\bar{\gamma}}$ and $k^i$ represent overall continuous momenta 
along coordinates $\gamma$, $\bar{\gamma}$ and $X^i$, respectively. 

The OPEs of the vertex operator $V_0$ with the fields $\beta$ and $\gamma$ are  
\begin{eqnarray}
\beta(z_1) ~V_0 (k^\gamma, k^{\bar{\gamma}}, k^i; z_2, \bar{z}_2) 
&\sim& -\frac{ik^\gamma}{z_{12}} ~V_0 (k^\gamma, k^{\bar{\gamma}}, k^i; z_2, \bar{z}_2), 
\label{zerob}\\
\gamma(z_1) ~V_0 (k^\gamma, k^{\bar{\gamma}}, k^i; z_2, \bar{z}_2) 
&\sim& ip' \ln (z_{12}) ~V_0 (k^\gamma, k^{\bar{\gamma}}, k^i; z_2, \bar{z}_2).
\end{eqnarray}
From the first equation we can read off the fact that the state corresponding to the vertex operator $V_0(k^\gamma, k^{\bar{\gamma}}, k^i; z, \bar{z})$ is an eigenstate of the zero-mode $\beta_0$ of the field $\beta(z)$, and the eigenvalue is $ik^\gamma$. Similarly, the vertex operator has eigenvalue $ik^{\bar{\gamma}}$ with respect to the zero-mode of the field $\bar{\beta} (\bar{z})$.

We can calculate the conformal weight of the vertex operator 
$V_{0}(k^\gamma, k^{\bar{\gamma}}, k^i; z, \bar{z})$ using OPE with the energy momentum tensor 
\begin{eqnarray}
T^{matter} (z_1) ~V_{0} (k^\gamma, k^{\bar{\gamma}}, k^i; z_2, \bar{z}_2)  ~\sim~ 
\frac{\left(\frac{\alpha'}{4} k^ik_i - k^\gamma p' \right)}{z_{12}^2} 
~V_0 (k^\gamma, k^{\bar{\gamma}}, k^i; z, \bar{z}) + \cdots \ , \label{holconformalweightforV}
\end{eqnarray}
where $T^{matter} (z_1) = T_{X}(z_1) + T_{\beta\gamma}(z_1)$. Thus the ground state vertex operator has following 
conformal weights for left and right moving sectors 
\begin{eqnarray}
(h_0, \tilde{h}_0)~ = ~\left(\frac{\alpha'}{4} k^ik_i - k^\gamma p' , 
~~\frac{\alpha'}{4} k^ik_i - k^{\bar{\gamma}} q'\right) \ .
\end{eqnarray}
To be a physical vertex operator we need to have $(h_0, \tilde{h}_0) = (1, 1)$. 
Thus there is a constraint to be imposed on all the physical operators, 
\begin{equation}
k^{\gamma} p' = k^{\bar{\gamma}} q' 
\label{lrcondition}
\end{equation}
Even though the commuting $\beta\gamma$ CFT is described by the first order formulation, 
it actually gives us zero mode contributions through an integral form of the field $\beta$ 
in the exponent of the vertex operator. These are related to the overall motion of these 
non-relativistic string states after we change fields $\gamma$ and $\bar{\gamma}$ 
into time and space in target spacetime.

As we already constructed the ground state vertex operator, 
it is relatively easy to construct first excited vertex operators, which 
can be written\footnote
{Covariant notation here is only 
for convenience and compactness.}
\begin{eqnarray}
V_e(k^\gamma, k^{\bar{\gamma}}, k^i; z, \bar{z}) = 
g : e_{MN} \partial X^{M} \bar{\partial} X^{N} e^{ik^\gamma \gamma + ik^{\bar{\gamma}} \bar{\gamma} 
- ip' \int^{z} \beta - iq' \int^{\bar{z}} \bar{b} + ik^i \cdot X_i} : , 
\label{vertexoperator2}
\end{eqnarray}
with 
\begin{eqnarray}
\partial X^M = \left(\partial \gamma, ~ \beta, ~ 
\left(2/\alpha'\right)^{1/2} \partial X^i \right), \\
\bar{\partial} X^N = \left(\bar{\partial} \bar{\gamma}, ~ \bar{\beta}, 
~ \left(2/\alpha'\right)^{1/2} \bar{\partial} X^j \right).
\end{eqnarray}
Note that $\partial X^M$ and $\bar{\partial} X^N$ have conformal weights 
$(1, 0)$ and $(0, 1)$, respectively. So the overall conformal 
weights of the first excited vertex operators are 
\begin{eqnarray}
(h_e, \tilde{h}_e)~ = ~\left(\frac{\alpha'}{4} k^ik_i - k^\gamma p' +1 , 
~~\frac{\alpha'}{4} k^ik_i - k^{\bar{\gamma}} q' + 1\right).
\end{eqnarray}
For this vertex to be physical we need to impose conditions 
$(h_e, \tilde{h}_e) = (1, 1)$. This gives us a familiar non-relativistic 
dispersion relation as we will see later. Higher excited vertex operators 
can be also constructed in a similar way.

\subsection{OCQ}

In the old covariant quantization scheme we can just forget about the ghosts, and 
can restrict to a smaller physical Hilbert space so that the missing equations of motion 
of the energy momentum tensor hold for matrix elements. Resulting requirements 
for the physical states $|  \psi \rangle$ 
can be written as\cite{pol}
\begin{eqnarray}
\left(L_{n}^{matter} + a \delta_{n,0} \right) | \psi\rangle &=& 0 ~~~~\hbox{for} ~~n \geq 0, \\
\langle\psi| L_{n}^{matter}| \psi'\rangle = \langle L_{-n}^{matter}\psi| \psi'\rangle 
&=& 0 ~~~~~\hbox{for} ~~n < 0.
\end{eqnarray}
Here $a=-1$. This normal-ordering constant comes from the ghost contribution only 
because bosonic matter $\beta\gamma$ CFT with $\lambda = 1$ has the vanishing normal-ordering constant.  

\bigskip
\noindent{\it ground state}

The ground state is denoted by $| V_0 \rangle$,
\begin{eqnarray}
| V_0 \rangle &=& V_0 (k^\gamma, k^{\bar{\gamma}}, k^i; z = 0, \bar{z} = 0)~ | 0;0\rangle_{X^i} 
\otimes  | 0\rangle_{\beta\gamma\bar{\beta}\bar{\gamma}} \nonumber \\
&=& | 0; k^i \rangle_{X^i} \otimes  | 0; k^\gamma, k^{\bar{\gamma}}
\rangle_{\beta\gamma\bar{\beta}\bar{\gamma}}. 
\label{groundstate}
\end{eqnarray}
Norm of this ground state is given by 
\begin{eqnarray}
\langle V_0(k_2) |  V_0 (k_1) \rangle = (2\pi)^{26} ~  \delta^{24} (k_2^i - k_1^i) ~  
\delta (k_2^{\gamma} - k_1^\gamma) ~  \delta (k_2^{\bar{\gamma}} - k_1^{\bar{\gamma}}) \ .
\end{eqnarray}
The first physical state condition can be evaluated to give us the result
\begin{eqnarray}
(L_{0}^{X} + L_0^{\beta\gamma} -1) |  V_0 \rangle  = \left(\frac{\alpha'}{4} k^ik_i - k^\gamma p' -1\right)
|  V_0 \rangle = 0 \ . 
\end{eqnarray}
When we evaluate the $\beta\gamma$ CFT part, we can not just ignore the term involved with the 
zero modes in $L_0^{\beta\gamma}$ because 
a divergent contribution from the vertex operator renders it finite.\footnote
{This divergent contribution of the ground state vertex operator presents in 
the critical string theory and is nothing new.
In our case it is important to keep this divergent contribution in mind because of the first order 
form of the energy momentum tensor. We can check explicitly that this is consistent with the observations 
given in the previous sub-section. } 
For this ground state 
to be physical, the condition $ \frac{\alpha'}{4} k^ik_i - k^\gamma p' = 1$ should be imposed.
And there is a similar condition from antiholomorphic sector as 
$ \frac{\alpha'}{4} k^ik_i - k^{\bar{\gamma}} q' = 1$. This is consistent with the previous 
result (\ref{holconformalweightforV}) that the conformal weight of the vertex operator 
should be $(h, \tilde{h}) = (1, 1)$ to become a physical vertex operator. 

From these physical conditions we have nontrivial relations between the parameters 
in the vertex operator. The arbitrary looking parameters, $p'$ and $q'$, in the vertex 
operator are uniquely fixed by the overall momenta $k^\gamma$, $k^{\bar{\gamma}}$ and $k^i$ 
as follows 
\begin{eqnarray}
p' = \frac{1}{k^\gamma} \left(\frac{\alpha'}{4} k^ik_i - 1\right) \ , ~~~~ q' 
= \frac{1}{k^{\bar{\gamma}}} \left(\frac{\alpha'}{4} k^ik_i - 1\right) \ . \label{p'q'value}
\end{eqnarray}
Or we can view these equations as equations of $k^\gamma$ and $k^{\bar{\gamma}}$ in terms of $p'$, $q'$ and 
transverse momenta. This viewpoint will give us a more familiar 
dispersion relation of the non-relativistic string.

To have a non-relativistic dispersion relation, we need to take into account 
the selection time, $\Time = p\gamma + q\bar{\gamma} = \cos(\phi) \gamma 
+ \sin(\phi) \bar{\gamma}$. And we need to introduce another coordinate $\Space$ 
which can be written as 
$\Space = -q \gamma + p \bar{\gamma} =  -\sin(\phi) \gamma + \cos(\phi) \bar{\gamma} $. 
Then we have 
\begin{eqnarray}
\gamma = \cos(\phi) \Time - \sin(\phi) \Space \ , ~~~~~~~~~~ 
\bar{\gamma} = \sin(\phi) \Time + \cos(\phi) \Space \ .
\end{eqnarray}
From the action Eq. (\ref{bosonicaction1}) it is clear that $\beta$ and $\bar{\beta}$ are 
conjugate variables of $\gamma$ and $\bar{\gamma}$. So we can identify the energy and momentum 
as follows 
\begin{eqnarray}
\frac{\beta}{2\pi} = \cos(\phi) P_{\Time} - \sin(\phi) P_{\Space} \ , ~~~~ \frac{\bar{\beta}}{2\pi} 
= \sin(\phi) P_{\Time} + \cos(\phi) P_{\Space} \ .
\end{eqnarray}
If one picks up eigenvalues for both sides one will have the equations, $i\beta_0 
= \cos(\phi) p_{\Time} - \sin(\phi) p_{\Space}$ and $i\bar{\beta}_0 =  \sin(\phi) p_{\Time} 
+ \cos(\phi) p_{\Space}$. With the eigenvalues $\beta_0 = -ik^\gamma$ and $\bar{\beta}_0 
= -ik^{\bar{\gamma}}$ given in equation (\ref{zerob}) we can get 
\begin{eqnarray}
p_{\Time} &=& \cos(\phi) k^\gamma + \sin(\phi) k^{\bar{\gamma}} \ ,   \\
p_{\Space} &=& -\sin(\phi) k^\gamma + \cos(\phi) k^{\bar{\gamma}}  \ . 
\end{eqnarray}
So the energy can be decided by the momenta $k^\gamma$, $k^{\bar{\gamma}}$ and the 
selection parameter $\phi$. At first glance, these expressions look a little bit strange but 
we can re-express these results in terms of other parameters using the 
constraint given by equation $pp' = qq'$.\footnote
{This part is a little subtle. After changing fields from $\beta\gamma$ to $\Time$ and $\Space$, 
we need to examine the ground state vertex operator in terms of these new variables. 
Concentrating on zero modes we have 
\begin{equation}
\exp \Big(i p_\Time (\Time + i [pp' \log z + qq' \log \bar{z}]) 
+ i p_\Space (\Space - i [qp' \log z - pq' \log \bar{z}])\Big) \ .
\end{equation}
To be consistent we need to impose one of the following conditions. (A) $pp' = qq'$. 
Then $p_\Time = \frac{1}{pp'}\left(\frac{\alpha'}{4} k^ik_i - 1\right)$ and $p_\Space = 0$. 
(B) $qp' = -pq'$. Then $p_\Time = 0$ and $p_\Space = \frac{p'}{p}\left(\frac{\alpha'}{4} k^ik_i - 1\right)$.
Effectively the $\beta\gamma$ zero modes are mapped into one bosonic coordinate and its momentum. What about 
the other coordinate? I think it is hidden somewhere because if there is only $\Time$ coordinate, 
the central charge do not match after changing variables. It will be interesting to investigate further 
along this line. One can consider the coordinate $\Space$ as a compact coordinate by introducing 
winding modes for $\gamma$ and $\bar{\gamma}$ fields. It remains to be seen that this compact coordinate 
can make the hidden coordinate visible or not. In the main body we will follow with the first condition. }

Then the results are 
\begin{equation}
p_{\Time} = \frac{1}{p p'}\left(\frac{\alpha'}{4} k^ik_i - 1\right)\ , ~~~~~~ p_{\Space} = 0 \ .
\end{equation}
Surely we are familiar with the first expression. The Mass of the non-relativistic particle corresponding the 
this vertex operator is 
given by the selection sector parameter and the parameter $p'$ in the vertex operator. 
This energy spectrum has the same structure as that of Gomis and Ooguri\cite{Gomis:2000bd} 
if one identifies the mass of the particle with $\omega R/2$ 
\begin{equation}
\epsilon_{GO} = \frac{\omega R}{2\alpha'} + \frac{2}{\omega R}\left(\frac{\alpha'}{4} 
k^ik_i - 1\right) \ , \label{energyofgo1}
\end{equation}
where $\omega$ and $R$ are the winding number and the radius of the compact coordinate. Note that the 
first term in equation (\ref{energyofgo1}), which is constant, comes from the tension of the 
winding string and is related to the term proportional 
to $\partial \beta \bar{\partial} \bar{\beta}$, which we ignored. 
They need a compact coordinate to have non zero result 
on the energy spectrum and energy is crucially related to the positive winding 
number of the compact coordinate.

\bigskip
\noindent{\it First Excited states}
\smallskip

The first excited states can be constructed with the corresponding vertex operator 
\begin{eqnarray}
|  V_e \rangle &=& V_e(k^\gamma, k^{\bar{\gamma}}, k^i; z = 0, \bar{z} = 0)~ 
\left(  | 0;0 \rangle_{X^i} \otimes   | 0 \rangle_{\beta\gamma\bar{\beta}\bar{\gamma}}\right)  \\
&=& \left(e_{M\bar{N}} \alpha_{-1}^M \bar{\alpha}_{-1}^N \right) ~ 
\left( | 0; k^i  \rangle_{X^i} \otimes   | 0; k^\gamma, k^{\bar{\gamma}} 
\rangle_{\beta\gamma\bar{b}\bar{\gamma}} 
\right) \ . 
\end{eqnarray}
Where the index $M$ runs $\gamma$, $\beta$ and $i$, and the index $\bar{N}$ runs 
$\bar{\gamma}$, $\bar{\beta}$ and $\bar{i}$. Thus 
$e_M \alpha_{-1}^M = \gamma_{-1} e_\gamma + \beta_{-1} e_\beta + \alpha_{-1}^i e_i$.  
Again the notation is for convenience. 

These states should satisfy the physical state conditions $(L_0^m - 1) | V_e\rangle = 0$ 
and $(\bar{L}_0^m - 1) | V_e\rangle = 0$ and we have 
\begin{eqnarray}
\frac{\alpha'}{4} k^ik_i - k^\gamma p' = 0 \ , ~~~ \frac{\alpha'}{4} k^ik_i 
- k^{\bar{\gamma}} q' = 0 \ . \label{p'q'valueforexcitedstate}
\end{eqnarray}
And there are other nontrivial conditions we need to impose for these states, 
$L_1^m | V_e\rangle = 0$ for the holomorphic part and similarly $\bar{L}_1^m | V_e\rangle = 0$ 
for the antiholomorphic part. Concentrating on the holomorphic part we can get 
\begin{eqnarray}
&&L_1^m  |  V_e  \rangle = \left(k^M e_{M\bar{N}} \bar{\alpha}_{-1}^N \right) 
| V_e\rangle = 0 \ , \\
\hbox{where} 
&&k^M = \left(p',~~ k^\gamma , ~~ \left(\alpha' / 2\right)^{1/2} k^i \right) \ , 
~~~~~~e_M = \left(e_\gamma, ~~ e_\beta, ~~ e_i \right) \ . \label{kmvector1}
\end{eqnarray} 
Thus we have conditions, 
$k^M e_{M} = p' e_\gamma + k^{\gamma} e_\beta + (\alpha'/2 )^{1/2} k^i e_i = 0$, 
for general $\bar{N}$

There is a spurious state at this level
\begin{eqnarray}
&&L_{-1}^m  |  V_0  \rangle = l_M {\alpha}_{-1}^{ M} | V_0\rangle 
= (- k^\gamma \gamma_{-1} -p' \beta_{-1} +(\alpha'/2)^{1/2} 
k_i \alpha_{-1}^{i}) | V_0\rangle \ , \\
&&\hbox{where}~~ l_M = \left(-k^\gamma, ~~ -p', ~~ \left(\alpha' / 2\right)^{1/2} k^i \right) \ ,  
~~~~ \alpha_{-1}^M = (\gamma_{-1},~~ \beta_{-1},~~ \alpha_{-1}^{i}) \ . \label{lmvector1}
\end{eqnarray}
From the observation, $k^M (e_{M\bar{N}} + l_M \bar{\beta}_N + A_M \bar{l}_N) 
= 0$, and 
with the conditions, $k^M A_M =\bar{\beta}_N \bar{k}^N = 0$, we can check 
that this spurious state $L_{-1}^m |  V_0 \rangle$ is actually physical and null. 
To derive this result we use the fact $k^M l_M = 0$, which is a direct consequence 
of the first physical state condition (\ref{p'q'valueforexcitedstate}). These nontrivial 
equations reveal the equivalent relation 
\begin{eqnarray}
e_{M\bar{N}}~ \approx~ e_{M\bar{N}} + l_M \bar{\beta}_N + A_M \bar{l}_N  \ , 
~~~~\hbox{with}~~k^M A_M = \bar{\beta}_N \bar{k}^N = 0 \ . 
\end{eqnarray}
These equations for the equivalence relation are similar to the 
relativistic string theory. From this observation we can conclude that this non-relativistic 
string theory has same number of degrees of freedom with the corresponding 
relativistic string theory.

We can also analyze the energy dispersion relation for the first excited state. 
The derivation is almost same as the previous sub-section and the result is 
\begin{eqnarray}
E_e =  \frac{k^i k_i}{2M} \ , ~~\hbox{where}~~ 
M = \frac{2}{\alpha'} pp' =\frac{2}{\alpha'} \cos(\phi) p' \ .
\end{eqnarray}
This energy dispersion relation is exactly same as that of the known 
non-relativistic particles.

\subsection{BRST quantization}

We already quantize the non-relativistic bosonic string theory with the OCQ method. 
But it is still interesting to see the equivalence between the OCQ result and the BRST quantization result. 

We are given with the gauge-fixed action (\ref{bosonicaction1}) and we will closely follow \cite{pol}. 
The only difference comes from the $\beta\gamma$ matter sector 
and we have the BRST transformations concentrating on the holomorphic part 
\begin{eqnarray}
&&\delta_{B} X^i = i\epsilon c_g \partial X^i \ , ~~~
\delta_{B} \beta = i\epsilon c_g \partial \beta \ , ~~~~
\delta_{B} \gamma = i\epsilon c_g \partial \gamma \ , \nonumber \\
&&\delta_{B} b_g = i\epsilon \left(T^m + T^g \right) \ , ~~~~
\delta_{B} c_g = i\epsilon c_g \partial c_g \ . 
\end{eqnarray}
BRST transformations for the $\beta\gamma$ matter CFT are nothing but the conformal 
transformations. $T^{matter} = T_X + T_{\beta\gamma}$, where $T_{\beta\gamma}$ is given above. 
And the energy momentum tensor for the ghost part has a usual form 
$T^g = (\partial b_g) c_g  - \lambda_g \partial (b_g c_g)$. 
The BRST current and charge have the following forms 
\begin{eqnarray}
j_B &=& c_g T^m + \frac{1}{2} :c_g T^g: + \frac{3}{2} \partial^2 c_g  \ , \\
Q_B &=& \frac{1}{2\pi i} \oint dz j_B(z)
= \sum_{n = -\infty}^{\infty} \left(c_{gn} L_{-n}^{m} + c_{gn} L_{-n}^{g} \right) - c_{g0} \ .
\end{eqnarray}

The OPE between the BRST current and the energy momentum tensor are 
\begin{eqnarray}
j_B(z_1) j_B(z_2) &\sim& - \frac{c^m - 18}{2z_{12}^3} c_g \partial_2 c_g (z_2) 
- \frac{c^m - 18}{4z_{12}^2} c_g \partial_2^2 c_g (z_2) 
- \frac{c^m - 26}{12z_{12}} c_g \partial_2^3 c_g (z_2) \ , \\
T(z_1) j_B(z_2) &\sim& \frac{c^m - 26}{2z_{12}^4} c_g (z_2) 
+ \frac{1}{z_{12}^2} j_B(z_2) + \frac{1}{z_{12}} \partial_2 j_B(z_2) \ .  
\end{eqnarray}
From these equations we can read off the facts that the BRST charge is nilpotent and the BRST 
current is a conformal tensor only if the central charge of the matter sector $c^m = 26$.

The physical states of this non-relativistic string theory can be systematically constructed 
with the BRST cohomology using the nilpotent BRST operator $Q_B^2 = 0$. 
The inner product of the states can be defined with the identifications 
\begin{eqnarray}
\alpha_{m}^{i\dagger} ~=~ \alpha_{-m}^{i}, ~~~&\beta_m^{\dagger} ~=~ -\beta_{-m} \ ,& 
~~~ \gamma_m^{\dagger} ~=~ \gamma_{-m} \ ,  \nonumber \\
&b_{gm}^{\dagger} ~=~ b_{g-m} \ ,& ~~~ c_{gm}^{\dagger} ~=~  c_{g-m} \ .
\end{eqnarray}
The zero modes of the ghost fields force the inner product of the ground state to have the form
\begin{equation}
 \langle V_0'(k_2)  |  \bar{c}_{g0} c_{g0} |  V_0' (k_1) \rangle = i (2\pi)^{26}~
 \delta^{24} (k_2^i - k_1^i)~ \delta (k_2^{\gamma} - k_1^\gamma) ~
\delta (k_2^{\bar{\gamma}} - k_1^{\bar{\gamma}}) \ , \label{norm4}
\end{equation}
Where $ |  V_0'  \rangle =  |  V_0  \rangle \otimes  | 0 \rangle_g$ denotes product of the matter ground 
state with the ghost ground state. The $c_{g0}$ and $\bar{c}_{g0}$ insertions 
are necessary for nonzero results.

To get the physical ground state we need two conditions
\begin{eqnarray}
b_{g0}  |  V_0'  \rangle &=& 0 \ , \\ 
Q_B  |  V_0'  \rangle &=& 0 \ .
\end{eqnarray}
These imply $L_0 ~ |  V_0'  \rangle = \{Q_B, ~b_{g0}\} |  V_0'  \rangle = 0$. 
This condition with the antiholomorphic part actually give us the same mass 
shell conditions as the OCQ result given in equation (\ref{p'q'value}). 
There is no exact state at this level and the state $ |  V_0'  \rangle$ is a
 BRST cohomology class which is physical.

At the first excited level there are $(26+2)^2$ states. 
\begin{equation}
 |  V_e'  \rangle = \left(e_{\mu\bar{\nu}} \alpha_{-1}^{\mu} 
\bar{\alpha}_{-1}^{\nu}\right)  |  V_0'  \rangle \ ,
\end{equation}
where the index $\mu$ includes $i, \gamma, \beta, c_g$ and $b_g$, 
and the index $\bar{\nu}$ includes $i, \bar{\gamma}, \bar{\beta}, \bar{c}_g$ and $\bar{b}_g$. 
The norm of this excited state for the holomorphic part is given by   
\begin{eqnarray}
 \langle V_e' | \bar{c}_{g0} c_{g0} |  V_e'  \rangle 
= \left( e_M^* e_M + e_{bg}^* e_{bg} + e_{cg}^* e_{bg} \right) 
\langle V_0'  | \bar{c}_{g0} c_{g0}  |  V_0'  \rangle \ ,
\end{eqnarray}
where $e_M^* e_M = e_i^* e_i + e_\beta^* e_\gamma + e_\gamma^* e_\beta$, and the last expression 
$ \langle V_0'  |  \bar{c}_{g0} c_{g0}  |  V_0'  \rangle$ is already 
evaluated in equation (\ref{norm4}). The antiholomorphic counter part also has a similar form.

The BRST invariant condition for this first excited states can be analyzed 
independently for holomorphic and antiholomorphic sectors. Concentrating on the holomorphic sector, we have   
\begin{eqnarray}
0 &=& Q_B   |  V_e'  \rangle \nonumber \\
&=& \left( c_{g-1} \{\alpha_0^i e_i + k^\gamma e_\beta + p' e_\gamma \} 
+ e_{bg} \{ \alpha_0 \alpha_{-1} - p' \beta_{-1} 
- k^\gamma \gamma_{-1} \} \right)   |  V_g'  \rangle \nonumber \\
&=& \left(c_{g-1} k^M e_M + e_{gb} l_M \alpha^{M}_{-1}\right)   |  V_g'  \rangle \ ,
\end{eqnarray}
where the detail calculation can be done with the same procedure given in previous sub-section. 
$k^M$ and $l_M$ are given by the equations (\ref{kmvector1}) and (\ref{lmvector1}), 
respectively. Because $c_{g -1}$ and $\alpha^{M}_{-1}$ are creation operators, 
the BRST closed condition forces us to have 
\begin{equation}
k^M e_M = 0, ~~~~~~~~ e_{bg} = 0 \ .
\end{equation}
This is very similar to the relativistic case. 

There is an additional zero norm state created by $c_{g-1}$ and $l_M \alpha^{M}_{-1}$. 
A general state is of the same form as the first excited state with different coefficients 
\begin{eqnarray}
 |  \psi  \rangle = \left(e^{M'} \alpha_{M-1} + e_{bg}' b_{g -1} + e_{cg}' c_{g -1}\right)   
|  V_g'  \rangle \ .
\end{eqnarray}
The BRST exact state at this level is 
\begin{eqnarray}
Q_B  |  \psi  \rangle = \left(c_{g-1} k^M e_M' + e_{gb}' l_M \alpha^{M}_{-1}\right)   
|  V_g'  \rangle \ .
\end{eqnarray}
Thus the ghost state $c_{g-1}  |  V_0'  \rangle $ is BRST exact, 
while the ``polarization'' has the following equivalence relation 
\begin{eqnarray}
e^M ~\sim~ e^M + e_{bg}' l^M \ . 
\end{eqnarray}
This is the same result as the OCQ given in the previous section. 
So there are total $(24)^2$ states in the first excited level 
which is exactly same as the physical spectra of the relativistic closed string.

\subsection{Scattering Amplitudes with Vertex Operators}

In this section we calculate various scattering amplitudes 
with the ground state vertex operators following the paper\cite{Gomis:2000bd}. And we also show that 
the amplitudes factorize properly into non-relativistic string poles. 
The scattering amplitude can be written 
\begin{eqnarray}
 \langle \prod_{s = 1}^{n} V_{0s}(z_s)  \rangle 
= \int {\it D X^i}{\it D \gamma }{\it D \bar{\gamma }}{\it D \beta }{\it D \bar{\beta }}
~ e^{-S } \prod_{s = 1}^{n} V_0(k_s^\gamma , k_s^{\bar{\gamma }}, 
k_s^i; z_s, \bar{z}_s) \ , \label{4vertex}
\end{eqnarray}
where the action is given in equation (\ref{bosonicaction1}) and the vertex operator 
is given in Eq. (\ref{vertexoperator}). 

These scattering amplitudes can be calculated with the functional integral. 
The calculation for the quadratic $X^i$ part can be done with the Gaussian integral and the result 
is same as the one in \cite{Gomis:2000bd}\cite{pol}. The first order functional integral can be 
evaluate with the Lagrangian (\ref{bosonicaction1}) and the extremum is given by 
\begin{eqnarray}
\beta (z) = \sum_{s} - \frac{ik_s^\gamma }{z - z_s}, ~~~~~ 
\gamma (z) = \sum_{s}  ip_s' \ln(z-z_s) \ . \label{bcmin1}
\end{eqnarray}
Thus all the functional integrals can be trivially evaluated and there are also the 
contributions from the various OPEs between the vertex operators. The result can be given 
\begin{eqnarray}
 \langle \prod_{s = 1}^{n} V_{gs}(z_s)  \rangle = g^n \prod_{s \neq t} 
(z_s - z_t)^{-k_s^\gamma  p_t'}~ (\bar{z}_s - \bar{z}_t)^{-k_s^{\bar{\gamma }} q_t'}
|  z_s - z_t | ^{\frac{\alpha'}{2} k_s^i k_{ti}} \ . \label{4vertex2}
\end{eqnarray}
From this calculation it is straightforward to evaluate scattering amplitudes for 
any number of the ground state vertex operator $V_0$. 

To evaluate the string poles we need to go a little bit further. 
The second exponent can be expressed 
\begin{eqnarray}
k_s^{\bar{\gamma }} q_t' ~=~ \frac{\tan(\phi_s)}{\tan (\phi_t)} k_s^\gamma  p_t' 
~=~ k_s^\gamma  p_t'  \ ,
\end{eqnarray}
where we use the relations $k^\gamma p' = k^{\bar{\gamma}} q'$ given in (\ref{lrcondition}), 
constraint $pp' = qq'$ and $\phi_s = \phi_t$, 
which is guaranteed by the fact that any two particles in different selection sectors 
can not interact with each other with the Galilean transformations. 
Then the n-point vertex scattering amplitude (\ref{4vertex2}) can be simplified 
\begin{eqnarray}
 \langle \prod_{s = 1}^{n} V_{0s}(z_s)  \rangle &=& g^n \prod_{s < t} 
| z_s - z_t| ^{-2k_s^\gamma  p_t' - 2k_t^\gamma  p_s'+ \alpha'k_s^i k_{ti}} 
\label{absolutescattering}\\
&=& g^n \prod_{s < t} | z_s - z_t| ^{-2 \cos(\phi)(p_s' + p_t') (E_s + E_t) 
+ \frac{\alpha'}{2} 
\left(k_s^i + k_{t}^i \right)^2 - 4} \ , \nonumber
\end{eqnarray}
where we used the energy relation $E_s = \frac{1}{p_s' \cos(\phi)} 
\left(\frac{\alpha'}{4} k_s^ik_{si} - 1\right)$. 
So we have the following closed string poles 
\begin{equation}
E_s + E_t = \frac{\frac{\alpha'}{4} \left(k_s^i + k_{t}^i \right)^2 + m - 1}
{\cos(\phi)(p_s' + p_t')} \ .
\end{equation}
This is the closed string spectrum of non-relativistic string theory, 
and can be identified with the general formula given in \cite{Gomis:2000bd} 
with appropriate modifications. 

These scattering amplitudes also can be calculated with the operator formulation, and 
these two results are equivalent. Scattering amplitude for the excited states can be also 
calculated without difficulty following the procedure given here. 
We present a four vertex scattering amplitude for the excited states in the appendix.
Because this non-relativistic theory is less symmetric and is lack of covariant notation
the expression for the scattering amplitude is quite complicated .

\subsection{One Loop partition function}

It is important to check the modular invariance of this non-relativistic string theory, 
because a breakdown of 
the modular invariance may be thought of as a global anomaly of the reparametrization 
invariance in string theories. The one loop partition function with a modulus 
$\tau = \tau_1 + i \tau_2$ on torus can be given in the operator language as 
\begin{equation}
Z (\tau) = Tr \Big( \exp (2\pi i \tau_1 P - 2\pi \tau_2 H) \Big) \ ,
\end{equation} 
where $P = L_0 - \bar{L}_0$ and $H = L_0 + \bar{L}_0 -\frac{1}{24} (c+\bar{c})$ . 

There are three independent parts to be evaluated, the $X^i$ CFTs, the ghost $bc$ CFT and 
the matter $\beta\gamma$ CFT. Contributions from the $X^i$ CFTs and the ghost $bc$ CFT 
are well known\cite{pol} 
\begin{equation}
Z_X^{tot} =V_{24} Z_X^{24} =  V_{24} ~\Big( \left (4\pi^2 \alpha' \tau_2 \right)^{-1/2} 
| \eta(\tau) |^{-2} \Big)^{24} \ ,  ~~~~Z_{bc} = | \eta (\tau) |^4 \ ,
\end{equation} 
where Dedekind $\eta(\tau)$ function is given by 
$\eta(\tau) = q^{1/24} \prod_{n=1}^{\infty} (1 - q^n )$ with 
$q = e^{2\pi i \tau} $.

A new contribution from the matter $\beta\gamma$ CFT can be calculated similarly \footnote
{The delta function is inserted because any physical state needs this condition. 
When we evaluate the integral we only integrated over the range $0 \leq k^\gamma \leq \infty$ 
without losing a generality which can be achieved by adjusting $p'$.}
\begin{eqnarray}
Z_{\beta\gamma} &=& V_{\beta\gamma} (q\bar{q})^{-2/24} 
\int \frac{dk^\gamma dk^{\bar{\gamma}}}{(2\pi)^2} \delta (k^\gamma p' - k^{\bar{\gamma}} q')
q^{-k^\gamma p'} \bar{q}^{- k^{\bar{\gamma}} q'} 
\prod_{m= -\infty}^{\infty} 
\sum_{N_m , \bar{N}_m = 0}^{\infty} q^{m N_m} \bar{q}^{m \bar{N}_m} \nonumber \\
&=& \frac{V_{\beta\gamma}}{2p'q'} 
\Big( \left(4\pi^2 \alpha' \tau_2 \right)^{-1/2} | \eta (\tau) |^{-2} \Big)^2 \ .
\end{eqnarray}
The matter part of the partition function $Z_X \cdot Z_{\beta\gamma}$ 
is the same as that of the relativistic string theory upto the volume factor. 
And the total partition function is 
\begin{eqnarray}
Z_{total} = \frac{V_{24} V_{\beta\gamma}}{2p'q'} \int \frac{d^2 \tau}{16 \pi^2 \alpha' \tau_2^2}
\left(Z_X^{24} \right).
\end{eqnarray}
This shows that the non-relativistic string theory is modular invariant because the integrand 
$\frac{d^2 \tau}{\tau_2^2}$ is modular invariant and $Z_X = \left (4\pi^2 \alpha' \tau_2 \right)^{-1/2} 
| \eta(\tau) |^{-2}$ is itself modular invariant. 
The contributions from the excitations of the $\beta\gamma$ CFT in the partition function 
are cancelled by those of the $bc$ CFT.  
The contributions from the zero modes of the $\beta\gamma$ CFT actually contribute to 
the factor $\frac{1}{\tau_2}$ in order to ensure the modular invariance.

\subsection{open string}

For the open string, we can impose the boundary condition  
$T_{zz} = T_{\bar{z}\bar{z}}$ at Im$z = 0$. The fields need to have the conditions,
\begin{equation}
\partial X^i = \bar{\partial} X^i, ~~~ \beta = \bar{\beta}, ~~~ \gamma = \bar{\gamma} ~~~~~\hbox{at}~ z = 0. 
\end{equation}
Then the usual doubling trick to write the holomorphic and antiholomorphic fields in the 
upper half-plane in terms of holomorphic fields in the whole plane,
\begin{equation}
\beta(z) \equiv \bar{\beta}( \bar{z}'), ~~~ \gamma(z) \equiv \bar{\gamma}(\bar{z}'), 
~~~\hbox{Im}(z) \leq 0, ~~ z' = \bar{z}
\end{equation}

The holomorphic energy momentum tensor should match the antiholomorphic energy momentum tensor 
after Galilean boost transformation at $z = 0$ for the open string case. 
This imposes a condition $p=q$ for the selection sector, in which the non-relativistic open string 
theory is meaningful. Consequently there is a constraint $p' = q'$ for the open string vertex operator. 
We expect the rest of the quantization procedure is similar to the closed string case and is 
straightforward after the work of the closed string theory.

\section{Bosonic Theory with general Commuting $\beta\gamma$ CFT}

After quantizing the critical case with $\lambda = 1$ for the non-relativistic string, 
it is natural to ask if there are also meningful theories for the $\beta\gamma$ CFT with 
different conformal weights. Thus 
we start with the bosonic String theory action with a general commuting $\beta\gamma$ CFT 
in addition to the spatial $X^i$ CFT in conformal gauge.  
\begin{eqnarray}
S = \int \frac{d^2z}{2\pi} \left( \beta \bar{\partial} \gamma + \bar{\beta} \partial \bar{\gamma} 
+  \frac{1}{\alpha'} \partial X^i \bar{\partial} X_i + b_g \bar{\partial} \gamma_g 
+ \bar{b_g} \partial \bar{c_g}
\right) \ , \label{bosonicaction}
\end{eqnarray}
where $i$ runs from 1 to d for $X^i$ CFTs. The commuting matter $\beta\gamma$ CFT has 
the conformal weights $h(\beta) = \lambda$ and $h(\gamma) = 1 - \lambda$. 
The anticommuting ghost $bc$ CFT has the weight $h(b_g) = 2$ and $h(c_g) = -1$. 

For the general commuting $\beta\gamma$ CFT, the OPEs are given by\cite{fms}\cite{pol}  
\begin{equation}
\gamma(z_1) \beta(z_2) \sim  \frac{1}{z_{12}} \ , ~~~~~ \beta(z_1) \gamma(z_2) \sim -\frac{1}{z_{12}} \ .
\end{equation}
And the antiholomorphic fields satisfy the similar OPE. 
The mode expansion and hermiticiy property are 
\begin{eqnarray}
&&\gamma(z) = \sum_{n=-\infty}^{\infty} \frac{\gamma_n}{z^{n+ (1-\lambda)}}, ~~ \gamma_n^{\dagger} 
=  \gamma_{-n},    \\
&&\beta(z) = \sum_{n=-\infty}^{\infty} \frac{\beta_n}{z^{n+\lambda}}, ~~~~~~~ \beta_{n}^{\dagger} 
=  - \beta_{-n}. 
\end{eqnarray}
The holomorphic energy momentum tensor and its mode expansion are 
\begin{eqnarray}
T &=& (\partial \beta) \gamma - \lambda \partial(\beta\gamma) = \sum_{n=-\infty}^{\infty}  
\frac{L_n}{z^{n+2}}, \nonumber \\
L_n &=& \sum_{m=-\infty}^{\infty} (n\lambda - m) :\gamma_{n-m}\beta_{m}: + a \delta_{n,0},
\end{eqnarray}
where $a = -\frac{\lambda(\lambda - 1)}{2}$ for the commuting bosons with periodic 
boundary condition. For the interesting case $\lambda = 1$, this ordering constant vanishes 
as we saw in the previous section.

The central charge of the commuting $\beta\gamma$ CFT is $2(6 \lambda^2 - 6 \lambda +1)$. 
To have a consistent theory, central charge from the matter 
CFTs (the $X^i$ CFTs and the $\beta\gamma$ CFT) should cancel the central charge $-26$ from 
the reparametrization ghost $bc$ CFT. Thus we can have the following condition
\begin{equation}
d = 26 - 2(6 \lambda^2 - 6 \lambda +1)
\end{equation}
We present a table with different $\lambda$ and $d$ for the possible consistent theories. 

\bigskip 
\begin{table}[h]
\begin{tabular}{|l|c|c|c|c|c|c|c|c|c|}
\hline \hline
$\lambda$  
  & $\cdots$ & \hspace{.1 in}2\hspace{.1 in} &\hspace{.1 in} 3/2 \hspace{.1 in} &\hspace{.1 in}1
\hspace{.1 in}
 &\hspace{.1 in} 1/2 \hspace{.1 in}& \hspace{.1 in}0\hspace{.1 in} &\hspace{.1 in} -1/2\hspace{.1 in} 
& \hspace{.1 in}-1\hspace{.1 in} & $\cdots$ \\
\hline
$c_{\beta\gamma}$
  & $c_{\beta\gamma}$ $>$ 26 & 26 & 11 & 2 & -1 & 2 & 11 & 26 & $c_{\beta\gamma}$ $>$ 26 \\
\hline
$d$
 & $\cdots$ & 0 & 15 & 24 & 27 & 24 & 15 & 0 & $\cdots$  \\
\hline \hline
\end{tabular}
\caption{Table for the possible consistent bosonic string theories with the $\beta\gamma$ CFT with 
integer and half integer conformal weights. the conformal weight $\lambda$, 
the central charge and the number 
of the spatial dimensions of the target space are presented.
``$\cdots$'' represents the case with the ``negative number'' spatial dimensions.
}
\label{table:bosonictable1}
\end{table}
\bigskip

We present the immediate observations on these possible consistent non-relativistic string theories, 
which have Galilean symmetry mentioned in the previous section with the corresponding 
rotational invariance, $SO(d)$. First, there exist only a finite range of 
conformal weight $-1 \leq \lambda \leq 2$ to have a space and time interpretation for the theories 
with the general $\beta\gamma$ CFT. The maximum number of the spatial coordinates excluding the $\beta\gamma$ 
CFT are $27$ for the case of $\lambda = 1/2$. This is a rather special case, and is worthy 
of further investigation. 

Second, as we increase $\lambda$, the central charge of $\beta\gamma$ CFT decreases 
until $c_{\beta\gamma} = -1 $ for $\lambda = 1/2$ and then it increases. Usually two different 
conformal weights correspond to the same central charge and the same number of spatial coordinates. 
We already considered 
the critical case of $\lambda = 1$. The $\lambda = 0$ case seems to be exactly same as 
the critical case if we change 
the role of $\beta$ and $\gamma$, except possibly different space and time interpretations. 
 
Third, there are two possible theories with only the $\beta\gamma$ CFT and the $ b_g c_g $ CFT 
without spatial coordinates. We would like to call these as ``topological theories.'' 
These topological theories have only the zero modes corresponding to spacetime coordinates and momenta
without any excitations. 
There are some curious facts in these theories as we mention at the end of this sub-section. 

Fourth, it will be interesting to understand the theories with half integer conformal weight fields. 
For the first glance, it seems that there is no possible interpretation of time in target space because 
the fields with half integer conformal weight do not give us zero modes. But 
as is well known, it is required to have two different sectors, NS and R sectors, for the half integer 
conformal weight fields. These fields are very similar to the superconformal ghost fileds. 
Of course there we need be careful with the continuous zero modes in order to have spacetime 
interpretations. 

These non-relativistic string theories are very similar to the critical case except 
the zero modes of the $\beta\gamma$ CFT. 
Establishing zero modes will be a challenge for these theories, but we think these problem could be solved 
with spectral flow. Generally noncritical relativistic string theories are 
hard to understand. In some sense, these theories are ``noncritical''. If we understand these theories 
better, we may have more insights for the relativistic counterpart.

\bigskip
\noindent{\it unification of the first order CFTs}

There is a curiosity about a unification of the first order CFTs.
The bosonic $\beta\gamma$ CFT and the ghost $b_g c_g$ CFT can be unified in bigger 
multiplets, ``grand-multiplets,'' ${\bf v}$ and ${\bf w}$ with new 
field $\Theta_{gh}$ which carries the conformal weight, 
the U(1) ghost charge and the U(1) matter charge  
\begin{equation}
{\bf v} = \beta + \Theta_{gh} b_g, ~~~~~ 
{\bf w} = c_g + \Theta_{gh} \gamma . \nonumber
\end{equation}
If one investigates these grand multiplets a little further, one can read off 
that $\Theta_{gh}$ is anticommuting field with the conformal weight 
$\lambda - \lambda_g$, the matter U(1) charge $-1$ and the ghost number $1$. ${\bf v}$ is a
commuting multiplet with the conformal weight $\lambda -1/2$, 
the U(1) matter charge $-1$ and the ghost number $0$, 
whereas ${\bf w}$ is an anticommuting multiplet with the conformal weight $1-\lambda$, 
the U(1) matter charge $0$ and the ghost number $1$. 

With these observation we can rewrite 
the bosonic string action in a very simple form for holomorphic part
\begin{equation}
S_{{\bf v}{\bf w}} = \int \frac{d^2 z}{2\pi} d\Theta_{gh} ({\bf v} 
\bar{\partial} {\bf w}  ) = \int \frac{d^2 z }{2\pi}(\beta \bar{\partial} \gamma +
 b_g \bar{\partial} c_g). \nonumber
\end{equation}
Here we did not gauge the field $\Theta_{gh}$ as indicated in the usual derivative $\partial$. 

For the cases of $\lambda = 2$ and of $\lambda = -1$ mentioned as ``topological'' theories, 
only the $\beta\gamma$ CFT and the $bc$ CFT are present. 
It will be interesting to investigate these theories further. There are only the zero modes without 
any oscillator excitations.

\section{Conclusion}

In this paper we investigate a new possibility of string theory with Galilean symmetry 
as a global symmetry. The action has the matter $\beta\gamma$ CFT in addition to the usual 
$X$ CFTs and the ghost $b_g c_g$ CFT in the conformal gauge. 
This Galilean symmetry is realized by the combinations of simplest theories, $\beta\gamma$ CFT and $X$ CFTs.
We quantize this theory in an elementary fashion. 
We would like to say that this theory has the full fledged form of a string theory and is at the same level 
as the perturbative relativistic string theory described 
by the Polyakov action. We think there are many different avenues ahead to be investigated.   

Why do we consider the non-relativistic string theories? 
First, the complete fundamental M-theory and string theory formulations are not available yet.
Recently a nonperturbative formulation of noncritical M-thery using a non-relativistic setup has been 
put forward\cite{Horava:2005tt}.
It seems to us that it is important to investigate other possibilities\footnote{Lorentz violating effects 
in string theory is very interesting and there are a lot of efforts to investigate them.
For example, see {\it e.g.} \cite{Ganor:2006ub}.}
 and to construct explicit examples, 
which will provide insights into the issues to be solved in the relativistic string theory. 
In this spirit, this non-relativistic string theory provides an example of a full bosonic string 
theory similar to the relativistic bosonic string theory in many aspects. 
In this nonrelativistic setup, it is possible to ask questions related to the nature of string theory 
even without the complication of gravity.
Second, it is possible that diffeomorphism invariance and Lorentz invariance can emerge at 
a special point of the moduli space of less symmetric theories such as the non-relativistic theories 
({\it e.g.}, \cite{cn1983}\cite{vol2001}). 
Third, even though there are many string theories with broken Lorentz symmetries such as lightlike 
Linear Dilaton theory, there were not so many attempts to understand these theories 
with emphasis on their manifest symmetries from the starting point.\footnote
{We thank Professor Petr Ho\v{r}ava for pointing out this to us. 
This is actually one of the strong motivation for the current work.}  

Here is an another motivation. There is evidence which suggests that time is rather different from 
space in string theory. ``Emergent spaces'' in string theory is not hard to 
find in the literature, while ``emergent time'' poses many challenges for current understanding of quantum theory
({\it e.g.},\cite{Seiberg:2006wf}).  
As a specific example without the complications of gravity, we can contrast 
``Noncommutative Open String theory(NCOS),'' which is a string theory 
with all the massive excitations of the open string in it, 
and ``Noncommutative Yang-Mills theory(NCYM),'' which is a field theory. 
These two theories are related to consistent low energy limits of the relativistic string theory 
with D-branes in the presence of NSNS B-field with an electric and a magnetic component, respectively. 
This non-relativistic string theory is different from the relativistic string theory by 
having the $\beta\gamma$ fields instead of the $X^0$ and $X^1$ fields, and it provides an example 
which treats time and space in a different footing.   

We will conclude with a few comments and future directions. 
This theory seems to be very similar to the relativistic string theory in many aspects.  
The total degrees of freedom of this non-relativistic string theory are the same as those of the relativistic 
string theory, because all the excited degrees of freedom are the same. 
The consistency conditions of the two dimensional conformal field theory put strong constraints 
on the spectrum of the non-relativistic theory, which suppress the excitations of the $\beta\gamma$ CFT 
in the physical spectrum. 
Differences between the relativistic string and the non-relativistic string are related 
to the zero modes and their interpretations. 
The zero modes of the $\beta\gamma$ CFT are very important for the modular invariance. 
By the way, the conventional $\beta\gamma$ CFT has a U(1) symmetry. On the other hand, we want to have 
$\beta\gamma$ zero modes in the ground state vertex operator in order to have a space and time interpretation.
Thus this U(1) symmetry is broken. Related facts for the zero modes of the superconformal 
$\beta_g \gamma_g$ CFT were already considered in \cite{hmm1988} long ago. 

Even though we started with the $\beta\gamma$ CFT and $X$ CFTs, we are able to identify the time of the target space 
as a linear combination of $\gamma$ and $\bar{\gamma}$ through the explicit Galilean boost transformation. 
There is a paramater of selection sectors which is responsible for the 
nontrivial dispersion relation. While we change the variables from $\beta\gamma$ to time $\Time$ and space 
$\Space$ in target space, we encounter a peculiar fact that space $\Space$ is actually hidden and 
only time $\Time$ is visible, as explained in the main text. It will be interesting to check whether it is possible 
to make space $\Space$ be visible by including winding modes in this theory by 
compactifying the coordinate $\Space$ similarly to the case of Gomis and Ooguri.

There are other possibly consistent string theories with $\beta\gamma$ of different 
conformal weights. We have a viewpoint that these theories are the non-relativistic analogues 
of noncritical string theories. And the analysis seems rather involved because it is not 
clear how to put $\beta\gamma$ fields explicitly in the vertex operator for these zero modes 
to give the spacetime interpretations. We think that spectral flow can have a role for this analysis.

We are currently investigating the supersummetric generalizations of this non-relativistic string theories.
For the critical case with 8 spatial coordinates, we have anticommuting $bc$ CFT in addition to the usual 
$\psi^i, i = 2, \cdots, 9$ CFTs. They all have identical conformal weights $1/2$. Naively we can change 
fields from $bc$ to $\psi^0$ and $\psi^1$ and we could get an $SO(9,1)$ symmetry in the fermionic sector. 
But this is too naive and there is no transformation which maps from $bc$ to the other $\psi^i$s. 
Quantization seems to be straightforward at least for the critical case. 
And there also exist similar noncritical supersymmetric generalizations for the 
noncritical non-relativistic string theories. It turns out that there is an infinite range of possible consistent 
string theories for this non-relativistic setup. We hope to report this progress in a near future. 
It will be also interesting to investigate the connection 
between these theories and supercritical string theories.\cite{Aharony:2006ra, Hellerman:2006nx}

\section*{Acknowledgments}

It is a pleasure to thank Professor Ori Ganor for encouragements, for answering questions and, 
especially, for thorough reading, correcting and commenting on the draft. 
I am pleased to thank Professor Petr Ho\v{r}ava 
for introducing earlier works related to NRCS as a reading assignment for his class, 
for valuable comments and ideas along the way to develop this theory and related area and 
for comments on the draft. 
I also thank Professor Ashvin Vishwanath for discussions related to string theory 
in the non-relativistic setup. 
This work was supported in part by 
the Center of Theoretical Physics at UC Berkeley,
and in part by the Director, 
Office of Science,
Office of High Energy and Nuclear Physics, 
of the U.S. Department of
Energy under Contract DE-AC02-05CH11231.

\section*{Disclaimer}
This document was prepared as an account of work sponsored by the United States Government. While this document is believed to contain correct information, neither the United States Government nor any agency thereof, nor The Regents of the University of California, nor any of their employees, makes any warranty, express or implied, or assumes any legal responsibility for the accuracy, completeness, or usefulness of any information, apparatus, product, or process disclosed, or represents that its use would not infringe privately owned rights. Reference herein to any specific commercial product, process, or service by its trade name, trademark, manufacturer, or otherwise, does not necessarily constitute or imply its endorsement, recommendation, or favoring by the United States Government or any agency thereof, or The Regents of the University of California. The views and opinions of authors expressed herein do not necessarily state or reflect those of the United States Government or any agency thereof or The Regents of the University of California.

\section*{Appendix. Scattering Amplitude for excited vertex operators}

The scattering amplitude for the 4 excited state vertex operators can be expressed as 
\begin{equation}
\langle \prod_{s = 1}^{n}V_{es}(z_s)  \rangle = 
\int {\it D X^i}{\it D  \gamma }{\it D \bar{ \gamma }}{\it D  \beta }{\it D \bar{ \beta }}~ e^{-S_b} 
\prod_{s = 1}^{n} V_e(k_s^ \gamma , k_s^{\bar{ \gamma }}, k_s^i; z_s, \bar{z}_s), 
\end{equation}
where the action is again given in (\ref{bosonicaction1}) and the vertex operator 
is given in (\ref{vertexoperator2}). 

It will be very convenient to evaluate this functional integration using the technique given 
in \cite{pol}, which evaluate the exponential factors first at the minimum of each field. 
The minimum of each field is again given by equation (\ref{bcmin1}). With these minimum values 
we can change variables from $\partial x^M$ to $q^M$ which will be given explicitly below. 
And we write down everything in terms of these new variables. After that we have the following 
scattering amplitude 
\begin{equation}
g^n \prod_{s \neq t} (z_s - z_t)^{-k_s^ \gamma  p_t'}~ (\bar{z}_s - \bar{z}_t)^{-k_s^{\bar{ \gamma }} q_t'}~  
\big| z_s - z_t \big|^{\frac{\alpha'}{2} k_s^i k_{ti}}  \langle\prod_{s = 1}^{n} e_{M\bar{N}}^s  
[v_s^M + q_s^M  ]  [\bar{v}_s^N + \bar{q}_s^N  ]  \rangle,
\end{equation}
where 
\begin{eqnarray}
q_s^M &=& \partial X_s^M - v_s^M,~~~~~ \bar{q}_s^M = \bar{\partial} X_s^M - \bar{v}_s^M, \\
v_s^M &=& \sum_{t \neq s} \frac{p_s^M}{z_s - z_t} ~~\hbox{with}~~ p_s^M = \left(ik_s^ \gamma , ~ -ip_s',
 ~ -i \left(\alpha' / 2 \right)^{1/2} k_s^i\right), \\ 
\bar{v}_s^N &=& \sum_{t \neq s} \frac{\bar{p}_s^N}{\bar{z}_s - \bar{z}_t} ~~\hbox{with}~~ 
\bar{p}_s^M = \left(ik_s^{\bar{ \gamma }}, ~ -iq_s', ~ -i \left(\alpha' / 2 \right)^{1/2} k_s^j\right).
\end{eqnarray}

We still need to evaluate the expectation value of $q^M$ and $\bar{q}^M$. These are given by 
the sum over all contractions. The only non vanishing two contractions are given by 
\begin{eqnarray}
 \langle q_s^ \gamma  q_t^ \beta   \rangle =  \langle q_s^ \beta  q_t^ \gamma   \rangle = \frac{1 - \delta_{st}}{(z_s - z_t)^2}, &&~~~
 \langle\bar{q}_s^ \gamma  \bar{q}_t^ \beta   \rangle =  \langle\bar{q}_s^ \beta  \bar{q}_t^ \gamma   
\rangle = \frac{1 - \delta_{st}}
{(\bar{z}_s - \bar{z}_t)^2}, \\
 \langle q_s^i q_t^j  \rangle =  \langle q_s^j q_t^i  \rangle = \frac{(1 - \delta_{st}) 
\delta^{ij}}{(z_s - z_t)^2}, &&~~~
 \langle\bar{q}_s^i \bar{q}_t^j  \rangle =  \langle\bar{q}_s^j \bar{q}_t^i  
\rangle = \frac{(1 - \delta_{st}) 
\delta^{ij}}{(\bar{z}_s - \bar{z}_t)^2}.
\end{eqnarray}

Rather than evaluating this general scattering amplitude we can evaluate one specific example 
with ``polarization'' $e_{ \beta \bar{ \gamma }}^1 ~e_{ \gamma \bar{ \beta }}^2 
~e_{\partial X^i \bar{\partial} X^j}^3 
~e_{\partial X^k \bar{\partial} X^l}^4 $ for the 4 excited state vertex operators 
as an illustration. 
The expectation value can be evaluated straightforwardly and we have the following 
expression without ``polarization'' factor which will be introduced later 
\begin{eqnarray}
&& \frac{4}{\alpha^{'2}} \langle  [ v_1^ \beta  + q_1^ \beta   ]  [ v_2^ \gamma  + q_2^ \gamma   ]  
[ v_3^{\partial X^i} + q_3^{\partial X^i}  ]  
[ v_4^{\partial X^k} + q_4^{\partial X^k}  ]  [ \bar{v}_1^{\bar{ \gamma }} + \bar{q}_1^{\bar{ \gamma }}  ]  
[ \bar{v}_2^{\bar{ \beta }} + \bar{q}_2^{\bar{ \beta }}  ]  [ \bar{v}_3^{\bar{\partial} X^j} 
+ \bar{q}_3^{\bar{\partial} X^j}  ]  [ \bar{v}_4^{\bar{\partial} X^l} + \bar{q}_4^{\bar{\partial} X^l}  ]  
\rangle \nonumber \\
&&= \left( (\sum_{s\neq 1} \frac{-ip_s'}{z_1 - z_s}) (\sum_{t\neq 2} 
\frac{ik_t^ \gamma }{z_2 - z_t}) + \frac{1}{(z_1 - z_2)^2}  \right)  \left( (\sum_{u\neq 3} 
\frac{k_u^i}{z_3 - z_u}) (\sum_{v\neq 4} \frac{k_v^k}{z_4 - z_v}) 
+ \frac{\delta^{ik} \cdot 2/\alpha'}{(z_3 - z_4)^2} \right) \nonumber \\
&&\times  \left( (\sum_{s\neq 1} \frac{ik_s^{\bar{ \gamma }}}{\bar{z}_1 - \bar{z}_s}) 
(\sum_{t\neq 2} \frac{-iq_t'}{\bar{z}_2 - \bar{z}_t}) + \frac{1}{(\bar{z}_1 - \bar{z}_2)^2}  \right)   
\left( (\sum_{u\neq 3} \frac{k_u^i}{\bar{z}_3 - \bar{z}_u}) (\sum_{v\neq 4} 
\frac{k_v^k}{\bar{z}_4 - \bar{z}_v}) + \frac{\delta^{ik} \cdot 2/\alpha'}
{(\bar{z}_3 - \bar{z}_4)^2} \right) \nonumber \\
&&  \equiv \frac{4}{\alpha^{'2}}A_{1~2~3~4}^{ \beta \bar{ \gamma }  \gamma \bar{ \beta } i\bar{j}k\bar{l}} 
\ . \nonumber
\end{eqnarray}

Thus the scattering amplitude can be summarized as using the result given in (\ref{absolutescattering})
\begin{eqnarray}
&& \langle V_{e_{ \beta \bar{ \gamma }}}(z_1) ~V_{e_{  \gamma \bar{ \beta }}}(z_2) ~V_{e_{ij}}(z_3) ~V_{e_{kl}}(z_4)  \rangle \nonumber \\
&=&  g^4 \left( e_{ \beta \bar{ \gamma }}^1 ~e_{ \gamma \bar{ \beta }}^2 ~e_{\partial X^i \bar{\partial} X^j}^3 
~e_{\partial X^k \bar{\partial} X^l}^4 \cdot A_{1~2~3~4}^{ \beta \bar{ \gamma }  \gamma \bar{ \beta } i\bar{j}k
\bar{l}}\right)~ \prod_{s < t} |z_s - z_t|^{-2k_s^ \gamma  p_t' - 2k_t^ \gamma  p_s'+ \alpha' k_s^i k_{ti}}. 
\end{eqnarray}
The scattering amplitudes with both ground state vertex operators and first excited vertex 
operators can be evaluated in a straightforward manner with the procedure given in this appendix.

\end{document}